\documentclass[aps,pra,reprint,showpacs,twocolumn,longbibliography,nofootinbib]{revtex4-1}

\usepackage{graphicx}
\usepackage{dcolumn}
\usepackage{bm}
\usepackage{braket}
\usepackage{leftidx}
\usepackage{cancel}
\usepackage{amsmath}
\usepackage{color}
\usepackage[mathcal]{eucal}
\usepackage{float}
\usepackage[bottom]{footmisc}
\usepackage[english]{babel}

\begin{document}
\title{\bf Direct micro-structuring of Si(111) surfaces through nanosecond-laser Bessel beams}
\author{Erkan Demirci$^{\bf (1)}$}
\author{Elif Turkan Aksit Kaya$^{\bf (1)}$}
\author{Ramazan Sahin$^{\bf (2)}$}\thanks{rsahin@itu.edu.tr}

\affiliation{${\bf (1)}$ {Informatics and Information Security Research Center, Tubitak-Bilgem 41470, Turkey}}
\affiliation{${\bf (2)}$ {Vocational School of Technical Sciences, Akdeniz University, 07058 Antalya, Turkey}}

\begin{abstract}

We present here a nanosecond laser ablation of Si(111) wafer with diffraction-free (Bessel-$J_0$) beams. First, the Axicon a conical shaped optical element for beam shaping is characterized with a visible and infrared light sources. Then, ablation profiles are obtained with Bessel beams generated for $\alpha=1^{\circ}$ and $\alpha=20^{\circ}$ base angles, then they are characterized via a Scanning Electron, an Atomic Force and an Optical Interferometric Microscopes. Experimental results compared with theoretical predictions obtained by using Bessel functions. Results show that Bessel beams give possibility of straightforward micro-structuring of Si(111) samples because the only central spot of Bessel beams could damage the surface provided that the laser pulse energy becomes in energy-range we found in our experiments. Moreover, our results clearly indicate that reduced HAZ area due to thermal expansion in ns pulse regime are natural outcome of Bessel beams as opposed to Gauss beams of same spot size. Lastly, nanosecond pulses indicate two-ablation regimes regarding the pulse energy, the quality of structures can be enhanced by using high quality Gauss beams ($M^{2}\approx1$) and the size of fabricated structures can be much reduced by using larger base angle of Axicon (such as $\alpha=40^{\circ}$) in our scheme. 

\end{abstract}

\maketitle

\subsection{Introduction}

The laser treatment of surfaces has attracted a numerous attention not only by the research community \cite{Kabashin2010} but also among the industrial applications \cite{NOLL20081159} due to its straightforward results without touching the target surfaces. In addition, this method does not require any chemical pre-treatment \cite{Jeffery} necessity. However, both the resolution of fabricated defects or ablation profiles and the quality of generated structures depend on the pulse duration. Although there are different type of lasers to be involved in laser ablation processes (such as high-power continuous lasers, pulsed or ultra-short pulsed lasers etc.), ultra-short pulsed lasers among them provide better aspect ratio profiles and minimal side effects during micro-structuring of surfaces \cite{MOMMA199715, liu_1997, karabutuv_2006, pronko_1995,simon_1996}. It is generally assumed that there are two-different ablation mechanisms relevant to interaction duration-pulse regime in the literature. When the pulse duration of laser source is in the order of heat diffusion time, the thermal ablation regime occurs as the material is locally heated by the light source by increasing its local temperature before vaporization \cite{zhou_physical_2011, bauer_heat_pulse}. On the other hand, a very hot-plasma of electrons is created by the ultra-short pulsed laser source. When the energy of this hot-plasma of electrons can reach to enough level to break molecular bonding, local material removing without heating the sample can be achieved in this regime \cite{mazur_2006,trtica_2007,amer_femto_nano}.

Silicon (Si) as a semiconductor material with a high precise patterning of its surface is a very popular due to its electrical and optical properties in especially electronic and photonic applications not only as a fundamental platform \cite{zain_silicon, grist_silicon} but also as a perfect candidate \cite{yates_science}. Since the quality of devices and their response to optical or electrical excitations are directly based on the quality of the fabricated structures \cite{chong_2010,zuev_fabrication_2011,malinaskus_2016,kim_2013}. Although ultrafast-lasers (femtosecond or picosecond) provide straightforward solution for structuring of Si surfaces with a better precision \cite{her_1998}, their availability (since they are much more expensive than nanosecond lasers) and limitations in practice (as the focal diameter decreases, the length of depth of focus becomes very short to increase resolution in which very fine positioning of sample surface is mandatory) pushed the researchers to find better alternatives to eliminate limitations \cite{watanabe_2010,karimzadeh_2009,demirci_2019}.

Nowadays, since the standard method possesses some limitations (such as optical aberrations, necessity of very fine control of sample positioning or reducing the wavelength of laser source) in practice \cite{sahin_gold}, the laser processing of materials at nano- or micro-scale dimensions has been undergoing a revolution to employ of diffraction-free Bessel beams \cite{Courvoisier2013, sahin2016, sahin2015, sahin2014,sahin_graphene} in which a very small central spot with a longer depth of focus can be simultaneously achieved. Diffraction-free (Bessel) beams possesses an electrical field distribution in the transverse plane given by the $J_0$ is the zeroth order Bessel function of first kind as in Eq. \ref{bessel_E}.

\begin{equation}
E(r) = J_0 (k_r r)
\label{bessel_E}
\end{equation}

Here, $k_r$ and r are the transverse wavevector and radial coordinate, respectively. Although there are some alternatives (circular holes at sub-wavelength dimensions where a very large amount of incident energy of laser beam is stopped or Spatial Light Modulators (SLM)) for transforming of Gaussian beams to Bessel beams, Axicon (conical shaped lens) is the better candidate considering its simplicity and energy conversion. Therefore, theoretical assumptions are based on this geometry.

\begin{equation}
k_r = \frac{2 \pi}{\lambda} sin \beta
\label{wavevector}
\end{equation}

The transverse wavevector, $k_r$, can be written as a function of the exit beam half cone angle,  $\beta$, and given by Eq. \ref{beta} ($\alpha$ is the base angle of the Axicon and n is the refractive index of the Axicon material)
; for which $k_r$ is given by Eq. \ref{wavevector} where $\lambda$ used for the laser wavelength.

\begin{equation}
\beta = arcsin (n sin \alpha) - \alpha
\label{beta}
\end{equation}

Nevertheless, the majority of the studies found in literature were conducted with ultra-fast lasers. Here, we apply this method (diffraction-free beams) to pattern the surface of crystalline Si samples. This manuscript is as organized as follows; (i) first the properties of Bessel beams are presented. Then, the details about experimental procedure and results are given. Finally, all the results are concluded after discussion of experimental results in the light of theoretical assumptions.

\subsection{Experiments} \label{experiments}

A single kind large-scale Czochralski growth p-type (doped with Boron) Si(111) wafer is used in our ablation experiments. The diameter of Si(111) wafer with a thickness of $508 \pm 15$ $\mu$m is nearly 3-inch before it was divided into small parts for a series of ablation experiments under clean-room conditions. The electrical resistivity of samples are averagely measured as 15 $\Omega$/sq. All the samples are washed with acetone,
ethyl alcohol and deionized water in ultra-sonic cleaner for at least 20 min., respectively.   

The electro-optic Q-switch laser system (EKSPLA NL230) used in ablation experiments generates 5.5 ns laser pulses with a wavelength of 1064 nm (Nd:YAG) at its fundamental mode. The repetition rate and number of pulses of the laser system can be adjusted via its computer controlled software. The single pulse energy of the laser source can be tuned between 5 mJ up and 90 mJ (peak power is approximately 16.4 MW) via appropriate Q-switch delay adjustment. The measured pulse to pulse instability of the laser system is around 0.27 $\%$ at the maximum energy. The polarization of the laser is kept constant as horizontal during the experiments. The output of the laser system is \textit{Top Hat} in the near field, and scatters to a nearly \textit{Gaussian} in the far field where the ablation experiments were conducted with a beam diameter of 4.5 mm. The divergence of the beam was measured less than 0.8 mrad. The measured beam propagation ratio ($M^2$) with a beam profiler is less than 2.5.

\begin{figure}
\centering
\includegraphics [width=0.45\textwidth]{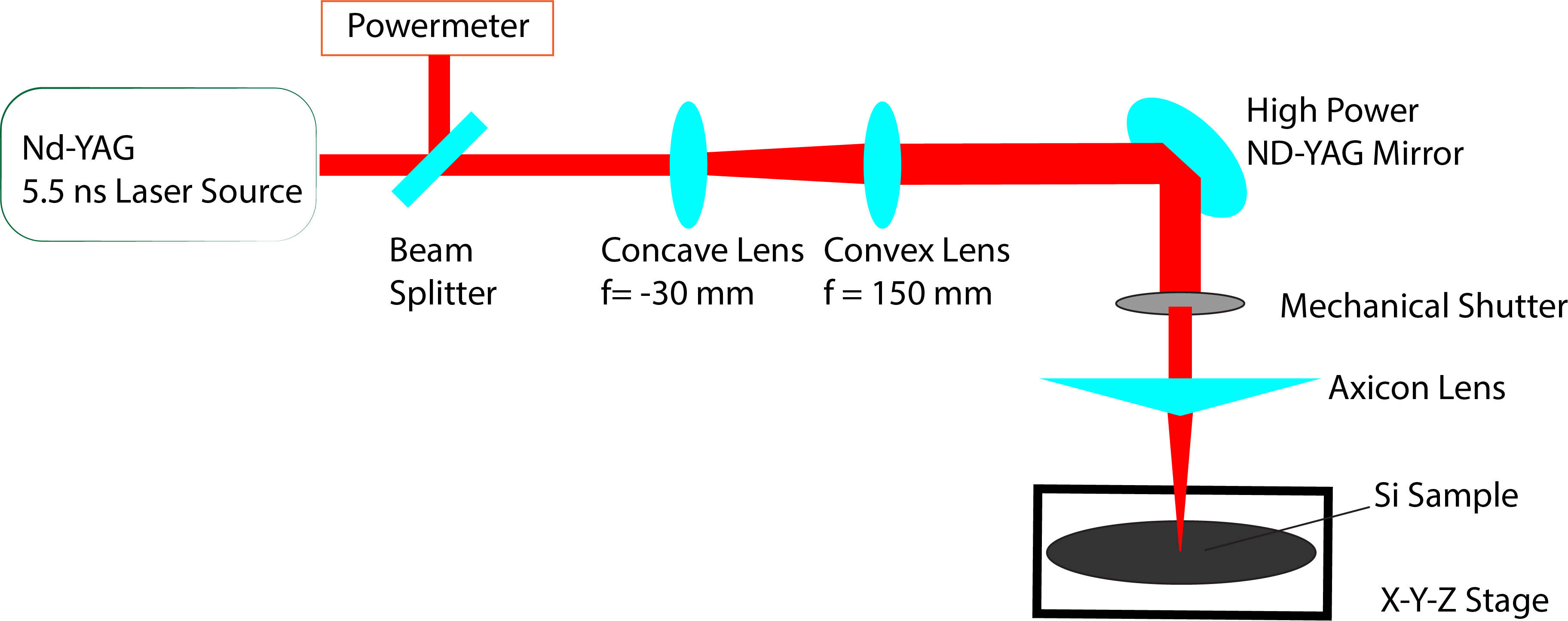}  
\caption{The experimental setup for Bessel beam processing of Si(111) samples}  
\vskip-0.3truecm
\label{fig1}
\end{figure} 

The experimental setup is illustrated in Fig.~\ref{fig1}. The beam is first expanded by using a plano-concave lens (Thorlabs LC4252, f=-30 mm) then collimated with a convex lens (Thorlabs LA4904, f=150 mm) to reduce laser-fluence on the sample surface and to enlarge the length of Bessel-zone. A high power ND-YAG laser mirror (Edmund optics 33-077) is directed to an Axicon lens (The two different Axicons are used in ablation experiments. The base angles of Axicon lenses made of fused silica are $1^{\circ}$ and $20^{\circ}$.) for beam shaping. Then, Bessel beams are directly focused on the sample surface where the Si(111) samples are placed on 3D translation stage. Before the axicon lens, a mechanical shutter with a diameter of 4.5 mm is used. The experiments are conducted in clean-room (ISO 8 class) conditions at room temperature (humidity is kept constant at $\%$50).

\begin{figure}
\centering
\includegraphics [width=0.4\textwidth]{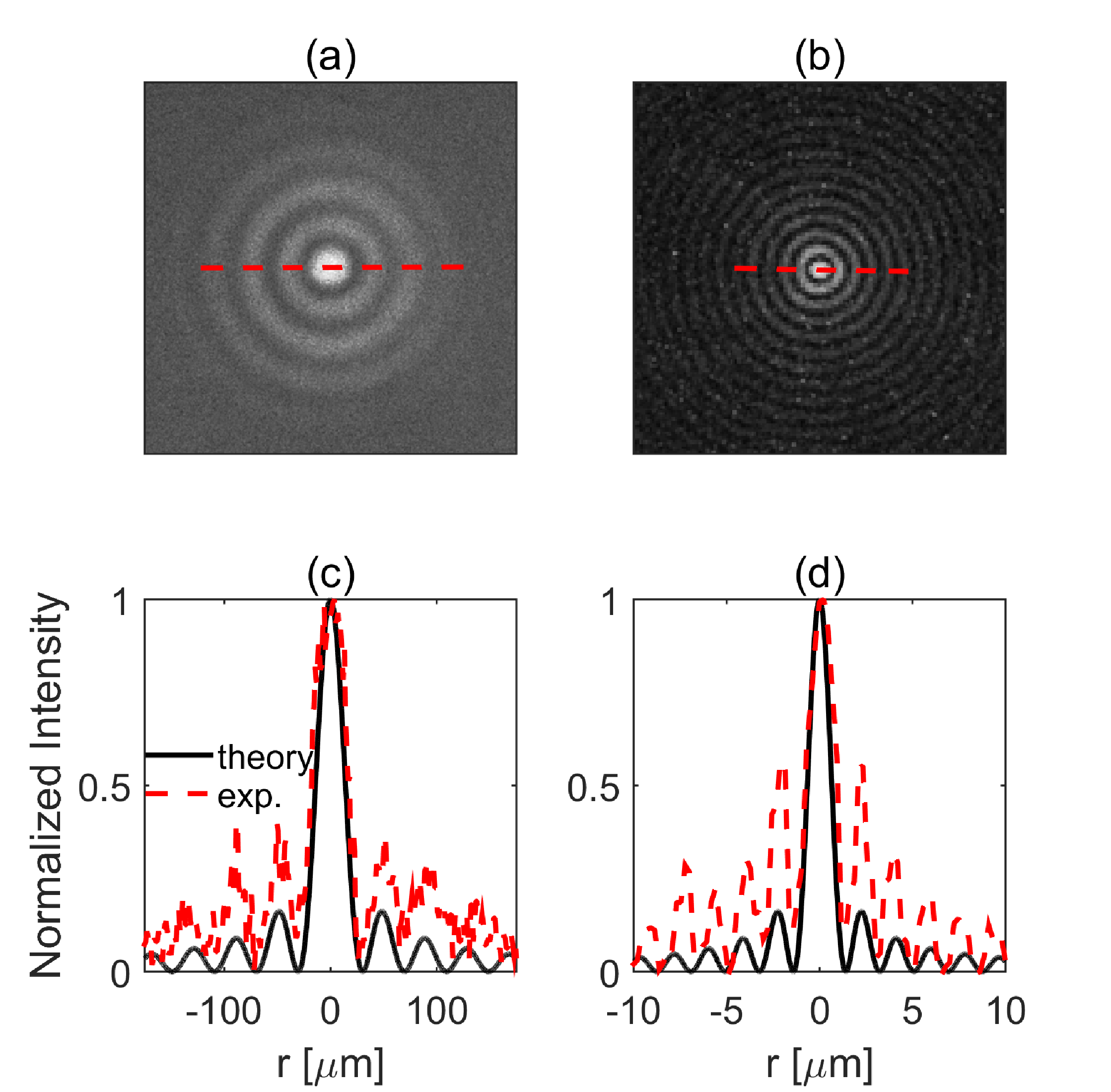}
\includegraphics [width=0.4\textwidth]{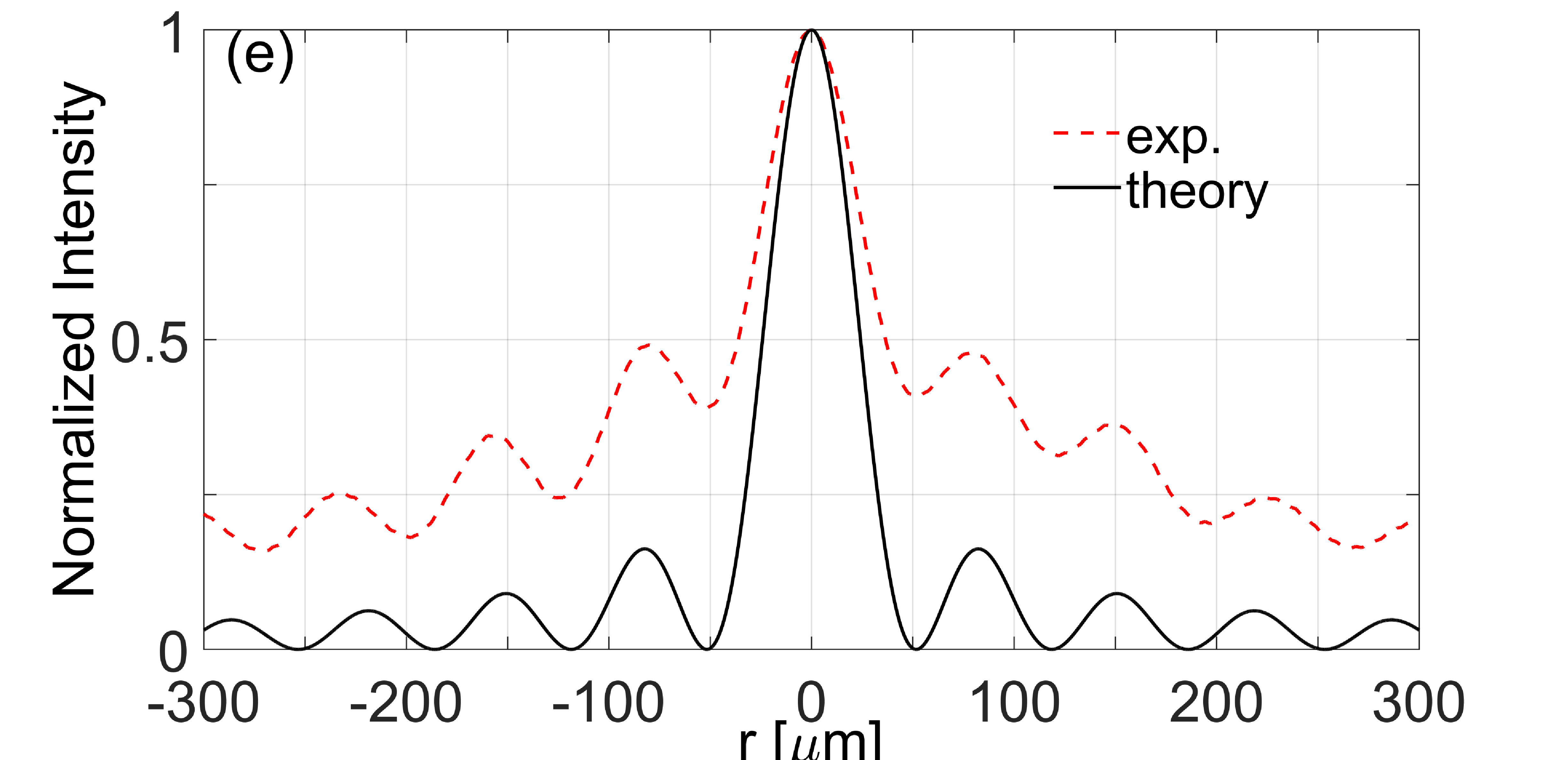}  
\caption{CCD camera images of Bessel beams for (a) $\alpha = 1^{\circ}$ and (b) $\alpha = 20^{\circ}$ base angles. (c) and (d) are their corresponding cross-sections (red dashed-line). He-Ne laser ($\lambda$=633 nm) is used for these experiments. (e) Experimentally obtained line-profile in the transverse plane after Axicon ($\alpha = 1^{\circ}$) for low-power continuous laser source operating at $\lambda$=1060 nm.}  
\vskip-0.3truecm
\label{fig2}
\end{figure}

The nanosecond-pulsed laser ($\lambda$=1064 nm) we use in our ablation experiments due to its peak power can damage the imaging system. Therefore, first, we characterize the Axicon lenses which would be used in beam-shaping and ablation experiments with visible light source for clarity. Fig.~\ref{fig2} indicates obtained images of beam profiles with a CCD camera. The cross-section through the red dashed-lines on the beam profiles are good agreement with the theoretical calculations calculated with Bessel functions ($J_0$) by using above equations. In addition, we also obtained Bessel beam (see Fig.~\ref{fig2} (e)) for continuous laser source of $\lambda$=1060 nm (in which this value is very close to wavelength of ns pulsed laser) with a beam profiler (Thorlabs BS209-IR) to be ensured that Bessel beams are successfully generated in the experimental setup.

\subsection{Results and Discussions}

Based on our theoretical calculations, the diameters of central spot of Bessel beams are 104 $\mu$m and 5 $\mu$m for $1^{\circ}$ and $20^{\circ}$ base angles respectively. In addition, the length of Bessel zone are found as 400 mm for $1^{\circ}$ and 20 mm for $20^{\circ}$.

Laser pulse fluence and number of laser pulses per shot play a significant role in the laser induced modifications of surfaces besides pulse duration. After Bessel beam generation, single pulse ablation experiments were immediately carried out on Si(111) sample. In the first part of the ablation experiments, the Axicon with a base angle of $1^{\circ}$ was used. The diameter of central lobe is calculated/ experimentally obtained as 104 $\mu$m.

\begin{figure}
\centering
\includegraphics [width=0.45\textwidth]{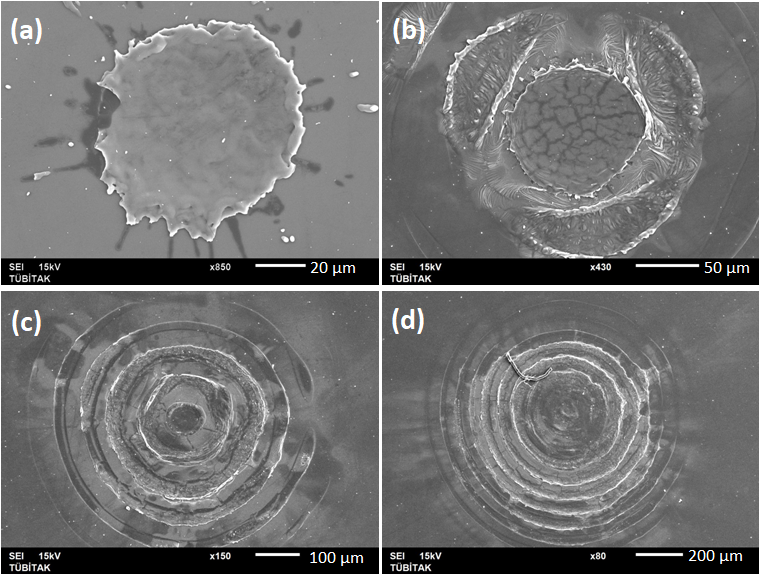}  
\caption{SEM images of ablated parts on Si(111) at single pulse regime with Bessel beams. The pulse energies are (a) 9.8 mJ, (b) 20 mJ, (c) 40 mJ, (d) 82 mJ.  }  
\vskip-0.3truecm
\label{fig3}
\end{figure}

\begin{figure}
\centering
\includegraphics [width=0.45\textwidth]{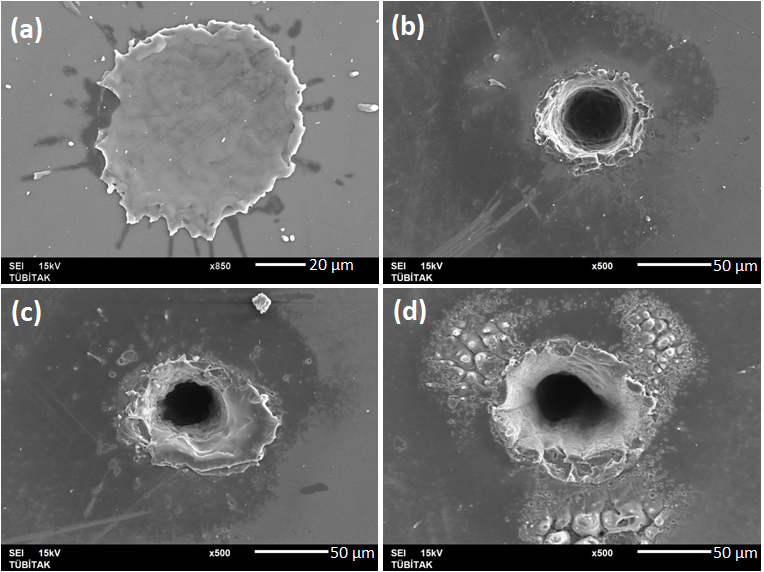}  
\caption{SEM images of ablated parts obtained from multi-pulse per shot experiments at a fixed pulse energy (9.8 mJ) in ambient conditions. The pulse numbers are (a) single-pulse, (b) 40 pulses, (c) 100 pulses, (d) 1000 pulses.}  
\vskip-0.3truecm
\label{fig4}
\end{figure}

As we gradually increase the laser pulse energy, we observe the ablation profile on the sample surface. Fig.~\ref{fig3} shows Scanning Electron Microscope (SEM) images of ablated parts on Si(111) sample. It is clear that ablation profiles approaches nearly Bessel beam profile (central spot surrounded by concentric rings) at larger laser pulse fluence which proves generation of Bessel beams after Axicon. Therefore, we found an energy window for ablation diameters comparable with the diameter of central spot of Bessel beams. Although we observed ablation patterns via SEM images even in single pulse regime (single central spot at relatively lower energies and central spot with much more complex surrounding rings at higher pulse energies, see Fig.~\ref{fig3}), the ablation depth are very small in those laser ablated structures. Therefore, we performed additional experiments if we achieved deeper ablation profiles. Fig.~\ref{fig4} shows SEM images of multi-pulses per shot ablation experiments at a constant pulse energy (9.8 mJ). 

We found in our experiments that the ablation depth of laser fabricated patterns could be much larger at relatively higher number of pulses per shot experiments. Since the laser pulse energy (9.8 mJ) is very low, only central spot of Bessel beam yields ablation as we observed in our SEM analysis. Moreover, the ablation diameter is smaller than the diameter of central spot (focal diameter) of Bessel beams which clearly indicates that the thermal expansion or Heat Affected Zone (HAZ) area is much smaller compared to laser fabrication with Gaussian beams \cite{demirci_2019}. In order to quantify the relation between laser pulse energy and ablation diameter, we conducted a systematic experiments for $\alpha=1^{\circ}$ by changing the laser pulse energy while observing ablation diameters.

\begin{figure}
\centering
\includegraphics [width=0.47\textwidth]{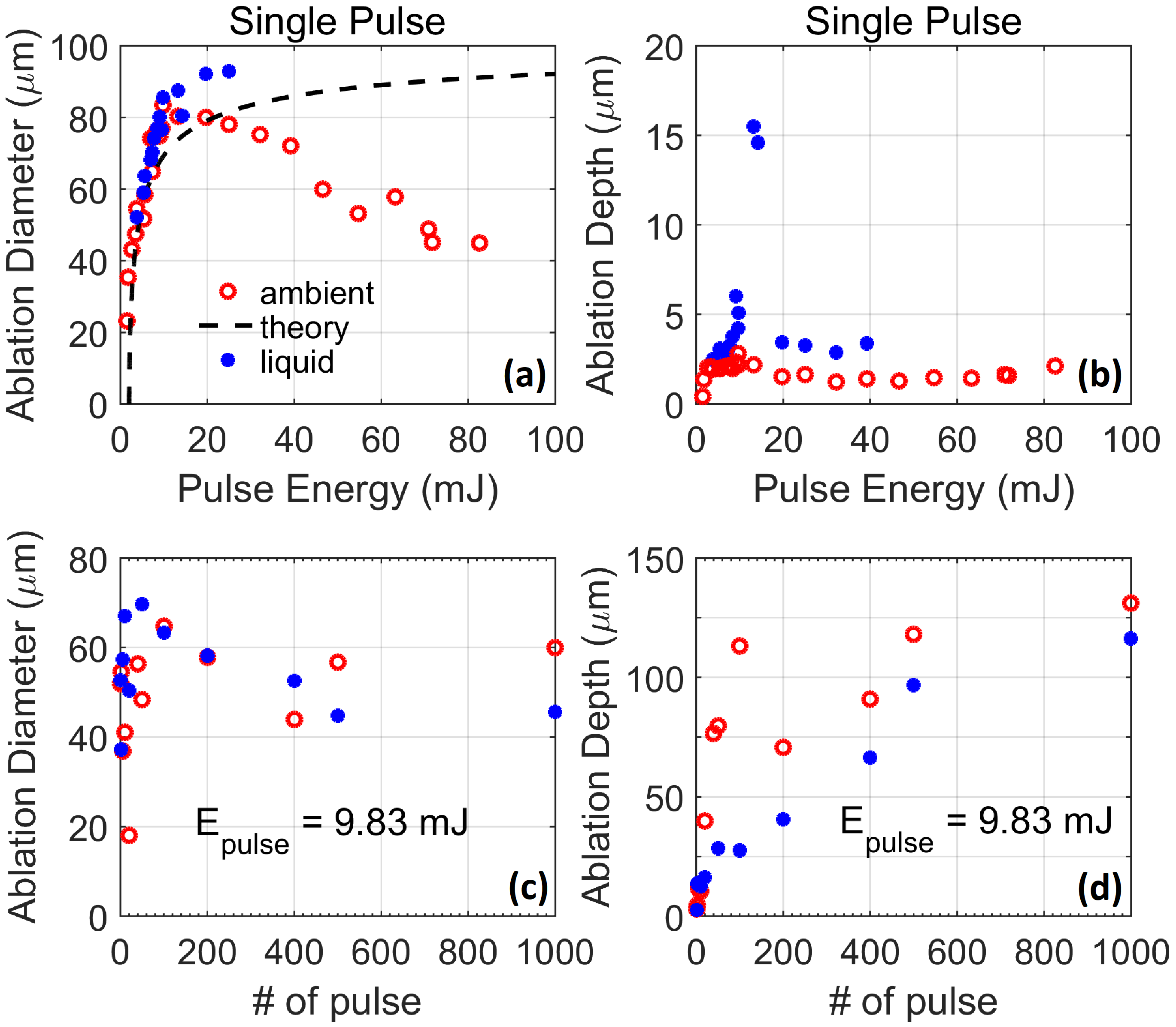}  
\caption{Effect of laser pulse energy and number of pulses values on the ablation diameter and ablation depth. Ablation diameter (a) and ablation depth (b) in single-pulse experiments. Black dashed-line in (a) is found by taking into account the mathematical expression of Bessel function ($J_0$) for $\alpha=1^{\circ}$. (c) and (d) are ablation diameter and ablation depth as a function of pulse number at a fixed pulse energy of 9.83 mJ, respectively. }  
\vskip-0.3truecm
\label{fig5}
\end{figure}

Results are shown in Fig.~\ref{fig5}. As the pulse energy increases, the resulting ablation diameter also increases as expected in low pulse energy regime where the only central spot of Bessel beam yields observable ablation. Since the surrounding rings can contribute to damage of sample in high pulse energy regime, ablation diameter does not obey the theoretical expectations as indicated in Fig.~\ref{fig5} (a). We think that those rings can modify the Si(111) surface even they would not ablation of it. Since we possess some experience from our previous work with Gauss beams that the liquid environment can cause reduced HAZ are, we performed additional experiments in deionized water environment. However, we could not observed a dramatic change in liquid environment. Therefore in the other parts of the experiments (for $\alpha=20^{\circ}$), we disregard its effect on the ablation and we do not employ ablation in liquid environment. Moreover, although Bessel beams propagates in a diffraction-free manner (this property is strongly dominant in transparent medium), a very short ablation depth in Si(111) values could be obtained in single-pulse ablation experiments (see Fig.~\ref{fig5} (b)). Since the only central spot of Bessel beams can contribute to ablation in low pulse energy regime (in broad energy range), we tried to increase ablation diameter in low pulse energy regime by adjusting the number of pulse per shot. Fig.~\ref{fig5} (c) indicates the results of SEM image analysis at a fixed pulse energy (9.83 mJ). The measured ablation diameter increases as the number of pulse value increases and then this approaches to a nearly constant (50 $\mu$m) value although the aspect ratio (the ratio of ablation depth and ablation diameter) linearly increases (see Fig.~\ref{fig5} (d)). 

\begin{figure}
\centering
\includegraphics [width=0.47\textwidth]{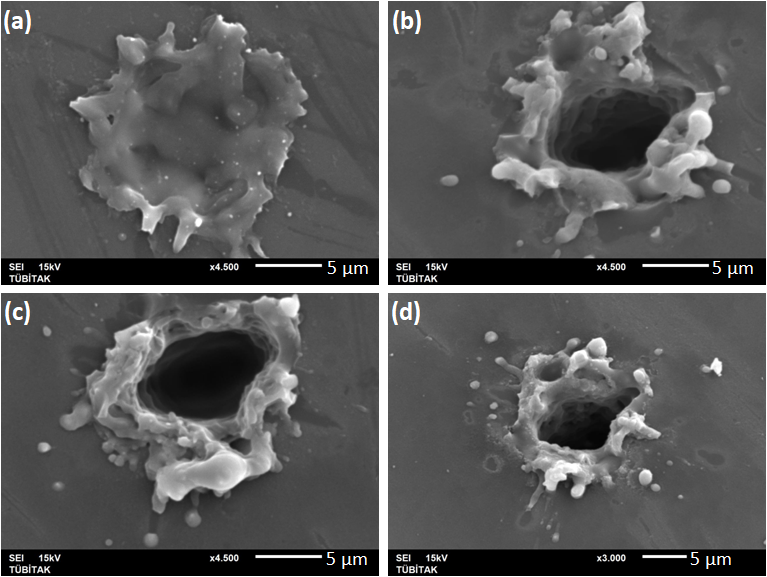}  
\caption{SEM images of ablated parts obtained from multi-pulse per shot experiments at a fixed pulse energy (15 mJ) in ambient conditions. The pulse numbers are (a) single-pulse, (b) 5 pulses, (c) 200 pulses, (d) 1000 pulses.}  
\vskip-0.3truecm
\label{fig6}
\end{figure}  

\begin{figure}
\centering
\includegraphics [width=0.47\textwidth]{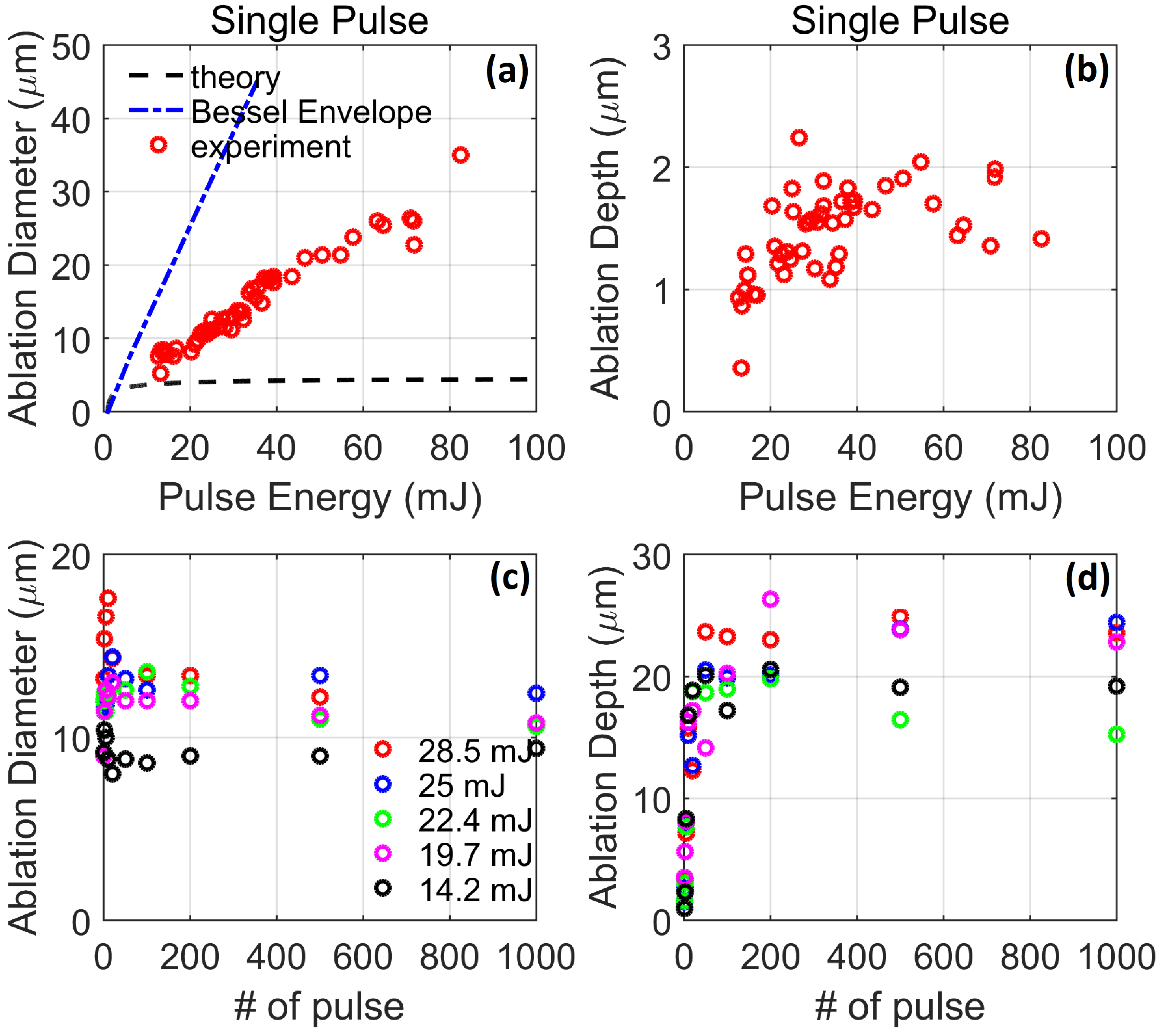}  
\caption{Effect of laser pulse energy and number of pulses values on the ablation diameter and ablation depth. (a) Ablation diameter and (b) ablation depth in sing-pulse experiments. Black dashed-line in (a) is found by taking into account the mathematical expression of Bessel function ($J_0$) and blue dashed-dotted line is found by using Bessel Envelope function for $\alpha=20^{\circ}$. (c) and (d) are ablation diameter and ablation depth as a function of pulse number at fixed pulse energies, respectively.}  
\vskip-0.3truecm
\label{fig7}
\end{figure}

We performed the same systematic experiments for $\alpha=20^{\circ}$ so that we can achieve a better resolution (minimal ablation diameter). SEM images of the first ablation experiments at a fixed pulse energy are shown in Fig.~\ref{fig6}. It is clearly seen that as the number of pulse increases the ablation profiles are deepen even though the ablation diameter stays nearly unchanged. For exact definition of ablation diameter as a function of pulse energy, we conducted a systematic pulse energy experiments while observing ablation diameter and ablation depth. Results are shown in Fig.~\ref{fig7}. The ablation diameters (red circles in Fig.~\ref{fig7} (a)) acquired as a function of pulse energy take the values between our theoretical predictions by taking into account ablation with only central spot of Bessel beam (black dashed-line in Fig.~\ref{fig7} (a)) and by taking into account the ablation with central spot and surrounding circles altogether (blue dashed-dotted line in Fig.~\ref{fig7} (a)). This assumption (blue dashed-dotted line) is calculated by using the envelope of Bessel function. These analysis show that ablation diameter in our experiments is much lower than the beam size which eliminates the inevitable results of using Gauss beams (due to heat diffusion in nanosecond pulse regime, ablation diameter is much higher than focal spot size). Even though we could not achieved a sub-micrometer resolution in our experiments, the ablation diameter of 5 $\mu$m could be generated and this value is well below the theoretical assumption (25 $\mu$m) with Gauss beams.  

The diameter of central spot is theoretically calculated as 5 $\mu$m for $\alpha=20^{\circ}$and Atomic Force Microscopy (AFM) results show that the ablation depth does not change dramatically in a very broad energy range (see Fig.~\ref{fig7} (b)). Moreover, ablation diameters at fixed energies (see Fig.~\ref{fig7} (c)) stays nearly unchanged (regardless of number of pulse) at low pulse energy regime although the ablation depth dramatically increases at first then approaches specific values (according to laser pulse energy) (see Fig.~\ref{fig7} (d)).    

\subsection{Conclusions}

As a conclusion, Bessel beams provides ablation diameters in which these diameters are much below the central spot of Bessel beams as opposed to Gaussian beams. Although ablation depth stays unchanged in the single pulse ablation regime as the laser pulse energy increases, a much higher aspect ratio structures could be fabricated in multi-pulse ablation regime. Moreover, Bessel beams propagate in a diffraction-free manner (means that the central spot size does not change in the propagation direction) but no dramatic changes of ablation diameter and depth have been observed in liquid environment. Although experimental values (ablation diameters) obey with theoretical assumptions especially in low pulse energy regime (for $\alpha=1^{\circ}$), those values deviates much from theoretical assumptions (where the size of central spot is taken into account). Therefore, we calculate predictions for ablation diameter (for $\alpha=20^{\circ}$) where Bessel envelope function is taken into account in our calculations. Since the quality of Bessel beam depends on the quality of Gauss beams (in which $M^2$ value is less than 2.5 in our setup), much lower values for the ablation diameter can be achieved by using a good Gaussian beams ($M^2 \approx 1$) and higher base angles of Axicon with an ultra sharp end (The bluntness of Axicon effects resulting beam shaping \cite{sahin_blunt_tip}.). These systematic experiments and results clearly indicate that a much higher resolution can be obtained via employing Bessel beams in surface processing.

\begin{acknowledgments}
R.Sahin thanks The Scientific Research Project Unit of Akdeniz University (Project Numbers: FAY-2017-2530, FYD-2017-3058, FYD-2017-3057, FYD-2019-4544). This work was also supported by TUBITAK (Project Number; 100082) and Ministry of Development, Republic of TURKEY (Project Number; 100139).
\end{acknowledgments}
%


%

\end{document}